\documentstyle{article}

\newcommand{\hsys}[1]{H_{sys}(#1)}
\newcommand{\hint}{H_I}
\newcommand{\hbath}{H_B}
\newcommand{\pot}[1]{V_{#1 \bmod N}}
\newcommand{\tr}{{\rm Tr}}
\newcommand{\re}{{\rm Re}}
\newcommand{\im}{{\rm Im}}
\newcommand{\comm}[2]{[#1,#2]}
\newcommand{\anticomm}[2]{[#1,#2]_+}

\begin{document}
\begin{titlepage}
\rightline{OUCMT-97-5}
\vfil
\leftline{\LARGE Quantum Ratchets\footnote{submitted to J. Phys. Soc. Jpn. }}
\vfil
\leftline{
  Satoshi {\sc Yukawa}\footnote{E-mail: 
    yukawa@ap.t.u-tokyo.ac.jp},
  Macoto {\sc Kikuchi}$^1$\footnote{E-mail:
    kikuchi@phys.sci.osaka-u.ac.jp} ,
  Gen {\sc Tatara}$^2$\footnote{E-mail:
    tatara@ess.sci.osaka-u.ac.jp},
  and Hiroshi {\sc Matsukawa}$^1$\footnote{E-mail:
    hiro@phys.wani.osaka-u.ac.jp}
}
\bigskip
\begin{flushleft}
{\it
  Department of Applied Physics,
  The University of Tokyo, Tokyo 113\\
  $^1$Department of Physics,
  Osaka University, Toyonaka 560\\
  $^2$Department of Earth and Space Science,
  Osaka University, Toyonaka 560
}
\end{flushleft}

\date{\today}

\vfil
\leftline{\bf Abstract.}
\medskip

\noindent

  The concept of thermal ratchets is extended to the system governed by
  quantum mechanics.  We study a tight-binding model with
  an asymmetric periodic potential contacting with a heat bath under
  an external oscillating field as a specific example of quantum ratchet.
  Dynamics of a density operator of this system is studied
  numerically by using the quantum Liouville equation.
  Finite net current is found in the non-equilibrium steady state.
  The direction of the current varies
  with parameters,
  in contrast with the classical thermal ratchets.
\vfil

\leftline{keywords:}
\noindent
ratchet, quantum nonequilibrium dynamics,  tight-binding model,
quantum tunneling, quantum Brownian motion

\end{titlepage}

Brownian particles under spatially asymmetric potential attract much
attention recently, which
has been shown to
produce nonzero net current under certain conditions\cite{AP92,Ma93}.
This type of the system is called {\it thermal} ratchet because
particles are
subject to
the thermal noise.  In the context of biology, thermal ratchets has
been discussed as a possible mechanism of biological
motors\cite{VO90,AB94,Ma94,SMB94}.  They also can be a mechanism of
actuators for micro- or nano-machines.  In the field of physics, it is studied
in connection with the noise induced
transportation\cite{DHR94,RSAP94,CM95,FBKL95,Mi95,ED96,Ja96,KCL96,ZBSH96}.

For nonzero net current to appear in thermal ratchets both the spatial
and time reversal symmetries need to be broken.  Otherwise, no finite
net current is expected, because of the Curie's principle\cite{Cu94},
which states that
asymmetric phenomena is caused by asymmetry (when no symmetry breaks
spontaneously).
Hence both (1) existence of an spatially asymmetric potential and (2)
a contact with a heat bath,
which breaks the time reversal symmetry of the system by
energy dissipation,
are required for the realization of thermal ratchets.  These two
requirements alone, however, can not produce finite net current due to
the second law of thermodynamics.  In order to make system
nonequilibrium,
and thus to allow steady current,
we require (3) existence of an additional noise or a time correlated
force.  A spatially unbiased noise or force is assumed hereafter,
because otherwise an occurrence of directional motion of the particle
is a trivial consequence of the
asymmetry.
We expect a directional particle motion in a non-equilibrium steady
state, if and only if above three conditions are fulfilled.

The models of thermal ratchets discussed so far in the literatures are
based on classical mechanics.  But there are situations in which
quantum effects such as quantum tunneling play important roles in the
particle motion; for instance, in small systems like mesoscopic
devices at low temperature.
Here we propose a new concept, {\it quantum ratchets}, which are ratchet
systems governed by quantum mechanics.
Quantum ratchets give rise to interesting
questions about nonequilibrium quantum dynamics
in the presence of interaction
with environments.  Moreover, by studying quantum ratchet systems, we
would be able to explore the new paradigm of quantum thermodynamics
such as a quantum engine or a quantum energy transducer.  In addition,
from the technological point of view, quantum ratchets have a
potential of being used as a new quantum electronic devices or quantum
motors for nano-machines in future.
In this letter, we propose one possible realization of quantum
ratchets and show that finite net current appear in that system, which
originates from purely quantum mechanical effects.

The model investigated in this letter is a tight-binding model.
Its Hamiltonian is given as follows:
\begin{eqnarray}
  H_{total} (t) & =  & H_{sys}(t) + \gamma_0 H_{int} + H_B, \\
  H_{sys}(t) & = & \displaystyle \sum_n \big\{ | n\rangle \langle n+1
  | + | n\rangle \langle n-1| + \big( \pot{n} +
  F_n(t) \big) |n\rangle\langle n| \big\}, \\
  H_B & = & \displaystyle \sum_\alpha \hbar \omega_\alpha
  \left(a_\alpha^\dagger a_\alpha + \frac{1}{2} \right),\\
  H_{int}
  & = & \displaystyle \sum_{n,\alpha} \big\{ g_{n,n+1} |
  n\rangle \langle n+1 | + g_{n-1,n} | n\rangle \langle n-1| \nonumber \\
 &+&\gamma_2 V_{n \bmod N} |n\rangle\langle n| \big\} \left(
    a_{\alpha}+a_{\alpha}^{\dagger} \right) \nonumber \\
  & \equiv &  H_I \displaystyle \sum_\alpha \left (
    a_{\alpha}+a_{\alpha}^{\dagger} \right).
\end{eqnarray}
$H_{total}(t)$ is Hamiltonian of the total system consisting of the
ratchet part and the heat bath.  The ratchet part is described by
$\hsys{t}$.
The localized state at site $n$
is represented by $ | n \rangle $.
$\pot{n}$ represents an asymmetric periodic potential with period $N$
and $F_n(t)$ an external time dependent field.
We measure energy in unit of the hopping coefficient.
The heat bath, which is described by $H_B$, is a collection of harmonic
oscillators,
each of which has a frequency $\omega_\alpha $ and is described by the
creation(annihilation) operator $ a_{\alpha}^\dagger (a_{\alpha}) $.
The ratchet part and the heat bath interact with each other with a
coupling constant $\gamma_0$ via $H_{int}$, which is factorized into
the following two parts: $H_I$ and $\sum_{\alpha}(a_\alpha^\dagger +
a_\alpha) $, which operate on the Hilbert space of the ratchet part
and the heat bath, respectively.  In $\hint$ the function $g_{n,m} $
is defined as $ 1 + \gamma_1 (\pot{n}+\pot{m})$ with the coupling
parameter $ \gamma_1 $; $\gamma_2 $ is another coupling parameter.
This interaction is a general form of real symmetric tridiagonal
interactions
up to the first order of $V_n$ (We omit 0-th order of the
diagonal part,
since it is a multiple of an identity operator and does not contribute
to the energy dissipation).
We note that the Hamiltonian $H_{total}(t)$ is Hermitian, because only the
diagonal part of $H_{total}(t)$ is spatially asymmetric.

Let us start with the quantum Liouville equation for the density
operator of the total system $\sigma(t)$,
\begin{equation}
  \frac{\partial \sigma(t)}{\partial t} = -
  \frac{i}{\hbar}\comm{H_{total}(t)}{\sigma(t)}.
\end{equation}
The density operator of the ratchet part $\rho(t) $ is obtained by $
\rho(t) = \tr_B \sigma(t) $, where $ \tr_B $ means the trace operation
with respect to the Hilbert space of the heat bath\cite{KTH91,Ga91},
Assuming that (1) the total density operator at initial time, $ t=0$,
is decomposed into the ratchet part and the heat bath part, i.e.,
$\sigma(0) = \rho(0)\rho_B$, (2) the heat bath is at its thermal
equilibrium at temperature $ T_{bath}$, so that $\rho_B = \exp \left(
  -\beta \hbath \right)/\tr_B \exp \left( -\beta \hbath \right), \beta
= 1/k_B T_{bath} $, and (3) the coupling constant $ \gamma_0 $ is
sufficiently small, we get the following equation for $\rho(t)$ after
standard calculations:
\begin{eqnarray}
  \frac{\partial \rho(t)}{\partial t}  &=& -\frac{i}{\hbar}
  \comm{\hsys{t}}{\rho(t)} \nonumber \\
  &-& \frac{\gamma_0^2}{\hbar^2}
  \int_0^t d s \Bigl(
  \comm{H_I}{H_I^{t,t-s}\rho(t-s)^{t,t-s}}\Phi(s) \nonumber \\
  &-&\comm{H_I}{\rho(t-s)^{t,t-s}
    H_I^{t,t-s}}\Phi(-s) \Bigr),\label{eq:bme}
\end{eqnarray}
where $\Phi(t) $ is the autocorrelation function of the operator $
\xi(t)$ $\equiv \exp\{ (i/\hbar) \hbath t \}$ $ \sum_\alpha (
a_{\alpha}+a_{\alpha}^{\dagger} ) \exp \{ -(i/\hbar) \hbath t \} $,
 i.e., $\Phi(t) = \tr_B \left\{ \rho_B \xi(t) \xi(0) \right\}$.  We
introduced the notation $A^{t,\tau}$ for an arbitrary operator $A$ as
follows:
\begin{equation}
  A^{t,\tau} = \overleftarrow{{\cal T}} \exp \left\{ -
    \frac{i}{\hbar} \int_\tau^t \hsys{t'} dt' \right\} A
  \overrightarrow{{\cal T}} \exp \left\{ \frac{i}{\hbar}
    \int_\tau^t \hsys{t'} dt' \right\},
\end{equation}
where $ \overleftarrow{{\cal T}},\overrightarrow{{\cal T}} $ indicate
time ordering product from the right to the left, and one from the
left to the right, respectively.  Equation~(\ref{eq:bme}) is exact up
to order $ \gamma_0^2$.

Since we have assumed the parameter $ \gamma_0 $ very small, the
characteristic time of the relaxation of $\rho (t) $ is very long.
Moreover, if the heat bath is sufficiently large, the function
$\Phi(t)$ vanishes quickly.  In such situation the integrand near $
s\sim 0 $ mainly contributes to the integral.  In addition, assuming
the external field changes more slowly than the characteristic time of
$\Phi(t)$, we can approximate eq.~(\ref{eq:bme}) as
\begin{equation}
  \frac{\partial \rho(t)}{\partial t} = -\frac{i}{\hbar}
  \comm{\hsys{t}}{\rho(t)} - \frac{\gamma_0^2}{\hbar^2}
  \Bigl( \comm{H_I}{\comm{{\cal K}(t)}{\rho(t)}} +
  i \comm{H_I}{\anticomm{{\cal H}(t)}{\rho(t)}}
  \Bigr)\label{eq:ame},
\end{equation}
where $\anticomm{A}{B} = AB+BA$.  Here we introduced new operators,
\begin{eqnarray}
  {\cal K}(t) & = & \int_0^\infty d s \exp \left\{ -
    \frac{i}{\hbar} \hsys{t} s \right\} H_I \exp
  \left\{ \frac{i}{\hbar} \hsys{t} s \right\} \re \Phi(s) \\
  {\cal H}(t) & = & \int_0^\infty d s \exp \left\{ -
    \frac{i}{\hbar} \hsys{t} s \right\} H_I \exp
  \left\{ \frac{i}{\hbar} \hsys{t} s \right\} \im \Phi(s).
\end{eqnarray}

In what follows, we study the dynamics of the system with an
asymmetric parabolic potential of period 5; i.e., $V_{0 \bmod 5} = 0
$, $V_{1 \bmod 5} = 1/9 $, $V_{2 \bmod 5} = 4/9 $, $V_{3 \bmod 5} = 1
$, and $V_{4 \bmod 5} = 1/9.$ Thus an unit cell consists of five
sites.  For modeling of the heat bath, we use the
spectral density of $\omega$ which produces
the standard Ohmic dissipation with the cutoff frequency
$1/\lambda$\cite{CL83,LCDFGZ87,We93}; then we can replace a summation
over modes by an integration, i.e., $\sum_\alpha \to \int_0^\infty d
\omega \omega \exp ( - \lambda \omega )$ in the evaluation of
$\Phi(s)$.  The external field is simply chosen as a periodically
on-off type ("flushing") field of the period $T$\cite{RSAP94,FBKL95};
\begin{equation}
  F_n (t) = \left\{
    \begin{array}{@{\,}ll}
      -\pot{n} & \mbox{for $ t \bmod T > T/2 $} \\ 0 & \mbox{for $
        t \bmod T \le T/2 $.}
    \end{array}
  \right.
\end{equation}
A current is defined through the conservation of the
probability; taking the diagonal element of the equation of
motion~(\ref{eq:ame}), we can define the local current $ J_n(t) $ at
the site $n$, by using the conservation relation for the probability $
\langle n | (\partial \rho(t)/\partial t) | n \rangle + J_{n+1}(t) -
J_n (t) = 0 $, as follows:
\begin{equation}
  J_n (t) = -\frac{2}{\hbar} \im \langle n | \rho(t) | n-1
  \rangle + \frac{2 \gamma_0^2}{\hbar^2} g_{n-1,n} \re
  \langle n | W(t) | n-1 \rangle,
\end{equation}
where $W(t) = \comm{{\cal K}(t)}{\rho(t)}+i \anticomm{{\cal
    H}(t)}{\rho(t)}$.  We integrate numerically each matrix element of
eq.~(\ref{eq:ame}) by the fourth-order Runge-Kutta
method\cite{PTVF92} (thus we solve complex simultaneous differential
equations).  We use the following parameters: $ \hbar = 1,$ $\gamma_0
= 0.01$,$\gamma_1 = 0.14$,$ \gamma_2 = 0.14$, and $ \lambda = 0.01$.
System size is taken to be 10 sites and the periodic boundary
condition is imposed, that is, the system consists of two unit cells.
All the numerical integrations are made with the time step of
Runge-Kutta method to be 0.0001.  The initial state of $\rho(t)$ is
prepared as the thermal equilibrium state of the inverse temperature
$\beta$.

In Fig.~(\ref{fig:tseries}) we show a typical time series of
the total current per unit cell $J \equiv \sum_{in~unitcell} J_n$.
Here
we set $\beta =0.8$ and $T=500$.  The first duration of 250 time
corresponds to the on-potential period in which the external force is
zero, and next one for 250 time is the off potential period in which
the external field cancels the asymmetric periodic potential.
Oscillation in the current are clearly seen in
Fig.~(\ref{fig:tseries}): in the on-potential periods, fast
oscillation appears but does not contribute to the time-averaged net
current.  Its typical time scale is the scale of $\hbar / \Delta E$,
where $ \Delta E$ is the typical eigenvalue difference
in the asymmetric potential.  On the other hand, much slower
fluctuation is observed in the off-potential period. This is
originated in the interaction Hamiltonian; its time scale is the order
of $1/\gamma_0$.  In the off-potential periods, we recognize a
synchronous appearance of the current, that is, every phase of the
oscillations in each off-potential period has an almost same value.
This synchronous oscillation of the current, which does not depend on
initial conditions and has a positive value for the present
parameters, mainly contributes to the net current when the long-time
average is taken.

In Figs.~(\ref{fig:phase}) we show the time-averaged current
$\langle J \rangle $, versus the inverse temperature $\beta$ (a), and
versus the period of the external field $T$ (b), respectively.  The
current varies with $ \beta $ and $ T $ and takes negative or positive
values according to the conditions.  This result is in clear contrast
with the classical thermal ratchets.  In the latter, the direction of
the current does not change with parameters.
In the present case the finite value of $J$ results from the
synchronous oscillation mentioned above.  The direction and the
magnitude of $J$ depend on the phase of the oscillation
at the instance the potential is turned off,
which depends on the parameters.

Contact with the heat bath is essential for the synchronous appearance
of the current.  In fact, such an synchronous oscillation disappear if
we turn off the interaction between the ratchet system and the heat
bath.  Figure~(\ref{fig:puretseries}) shows the current as the
function of time, as Fig.~(\ref{fig:tseries}),
but
in the absence of interaction with the heat bath.  In this situation,
the ratchet system obeys pure quantum dynamics.  The current still
appears in the off-potential periods, but its magnitude and direction
are random.  The reason is the following: in the on-potential periods,
the current oscillates according to the eigenvalue difference of the
Hamiltonian.  Since the period of external field and a oscillation
period in the on-potential period is generally incommensurate with
each other, the phase
of
the oscillation when the potential is turned off
fluctuates
almost randomly.

The mechanism of synchronous oscillation
for the case of finite $\gamma_{0}$
is considered as follows:
After large enough repeated time $n$ of the external field, the system
becomes non-equilibrium
steady state due to the dissipation.  Then the density matrix $\rho$
at each end of the period of the external field coincides with each
other.  It produces the synchronous oscillation.
It is to be noted that the above stationary motion does not mean the
stationary motion in each duration where the potential is constant.

To summarize, we proposed a new concept, {\it quantum ratchet}, and its one
possible realization.
Finite
net current is found in the non-equilibrium steady
state, which
originates from purely quantum mechanical effects.
The behavior of the current is
quite different from that in the classical one: in the quantum
ratchet, the current flows both direction which is
determined by the heat bath temperature and/or the period of the external
field, though, in the classical situation, it flows
to one-direction decided by asymmetry of potential and type
of external field.
The bidirectionality means that, in quantum ratchet devices, we can
control the electric current by changing parameters\cite{SGN97}.

The present model has been considered in the weak coupling situation.
Thus the relaxation time of the system becomes long. Because of this,
we have studied the system only in the non-adiabatic time scale.  If
we study in the adiabatic and strong coupling region, different
phenomena like crossover between classical and quantum regimes may be
observed. The present results depend strongly on the detail of the
model such as the potential shape, system size, and so on.
This indicates that
there is a possibility to produce the current more effectively by
changing details of the model.
For the study of the quantum thermodynamics of the rathcet system,
especially properties as a quantum engine, or a quantum energy
transducer, we have to study efficiency of energy\cite{Se96}.
Detailed study is now in progress.

The authors are grateful to S. Miyashita, T. Nagao, K.Saito for
valuable discussions.  This work is partly supported by a Research
Fellowships of the Japan Society for the Promotion of Science for
Young Scientists and also partly supported by
Grants-in-Aid from the Ministry of
Education, Science, Sports and Culture.

\newpage

\begin{figure}
\caption{
  Typical time series of the total current per unit cell: initial
  state is chosen as a thermal equilibrium state.  We take some
  parameters to be $\beta = 0.8 $ and $T=500$. The system in first
  duration for 250 time is in ``on-potential'' state. In next one for
  250 time it is in ``off-potential'' state. The total current $ J =
  \sum_{in~unitcell} J_n$ in the unit cell is calculated with the
  time interval 5.
  }
\label{fig:tseries}
\end{figure}

\begin{figure}
\caption{The average current $\langle J \rangle $ versus (a)
  the heat bath inverse temperature $\beta$ and (b) the period of the
  external field $T$.
  In a classical overdamp limit, we expect only negative current to
  appear.
  }
\label{fig:phase}
\end{figure}

\begin{figure}
\caption{
  Typical time series of the current under pure quantum dynamics, that
  is, in the absence of the interaction with the heat bath.  Initial
  condition of the system density matrix is taken to be ``thermal
  equilibrium'' state with inverse temperature $\beta = 0.8 $. All of
  the other conditions are same as Fig.~(\ref{fig:tseries}).  }
\label{fig:puretseries}
\end{figure}

\end{document}